\documentclass[runningheads]{llncs}
\usepackage{graphicx}
\usepackage{comment}
\usepackage{amsmath,amssymb} 
\usepackage{color}
\usepackage{multirow}
\usepackage{makecell}
\usepackage{threeparttable}
\usepackage{multicol}
\usepackage{subfigure}

\usepackage[colorlinks, citecolor=red, citecolor=green]{hyperref}
\usepackage{booktabs}
\usepackage{cite}
\usepackage{caption}

\begin{document}
\pagestyle{headings}
\mainmatter
\def\ECCVSubNumber{3375}  

\title{Interactive Multi-Dimension Modulation \\ with Dynamic Controllable Residual Learning\\for Image Restoration} 

\titlerunning{Multi-Dimension Modulation}
%

\author{Jingwen He\thanks{The first two authors are co-first authors. $\dagger$ Corresponding author}\inst{1,2} \and Chao Dong$^{\star}$\inst{1,2} \and Yu Qiao$\dagger$\inst{1,2}}
\authorrunning{Jingwen He et al.}
%
\institute{ShenZhen Key Lab of Computer Vision and Pattern Recognition, SIAT-SenseTime Joint Lab, Shenzhen Institutes of Advanced Technology, Chinese Academy of Sciences\\ \and
	SIAT Branch, Shenzhen Institute of Artificial Intelligence and Robotics for Society
	\email{\{jw.he, chao.dong, yu.qiao\}@siat.ac.cn}}
\maketitle

	
	\begin{abstract}

Interactive image restoration aims to generate restored images by adjusting a controlling coefficient which determines the restoration level. Previous works are restricted in modulating image with a single coefficient. However, real images always contain multiple types of degradation, which cannot be well determined by one coefficient. To make a step forward, this paper presents a new problem setup, called multi-dimension (MD) modulation, which aims at modulating output effects across multiple degradation types and levels. Compared with the previous single-dimension (SD) modulation, the MD is setup to handle multiple degradations adaptively and relief unbalanced learning problem in different degradations. We also propose a deep architecture - CResMD with newly introduced controllable residual connections for multi-dimension modulation. Specifically, we add a controlling variable on the conventional residual connection to allow a weighted summation of input and residual. The values of these weights are generated by another condition network. We further propose a new data sampling strategy based on beta distribution to balance different degradation types and levels. With corrupted image and degradation information as inputs, the network can output the corresponding restored image. By tweaking the condition vector, users can control the output effects in MD space at test time. Extensive experiments demonstrate that the proposed CResMD achieve excellent performance on both SD and MD modulation tasks. Code is available at \url{https://github.com/hejingwenhejingwen/CResMD}.
	\end{abstract}
	\begin{figure}[t]
		\centering
		\includegraphics[scale=0.32]{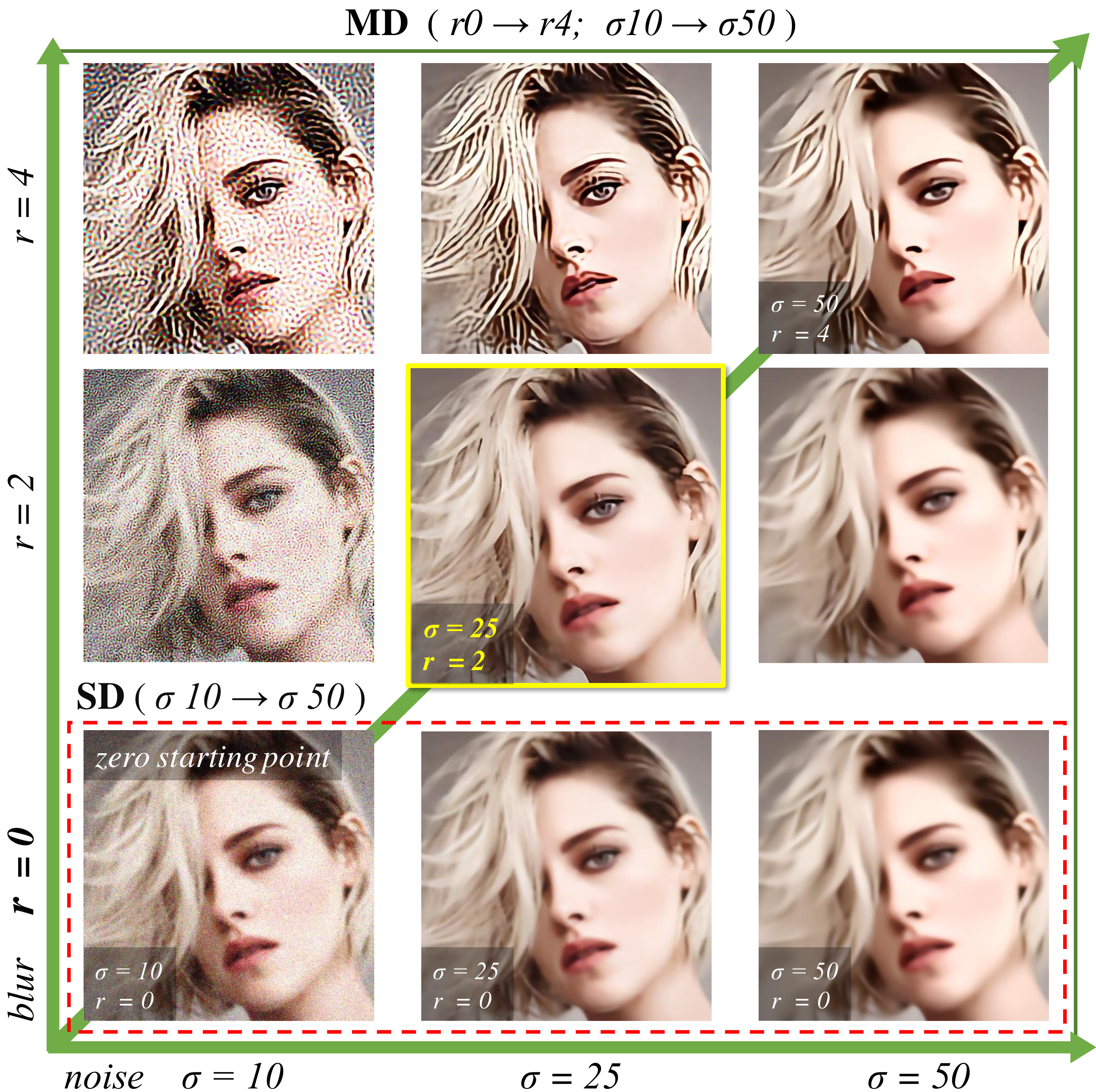}
		\caption{Two-dimension (2D) modulation for a corrupted image with blur $r2$+noise $\sigma25$. When the blur level is fixed to $r0$, we can only modulate the denoising effect ($\sigma10\rightarrow\sigma50$), which is a typical single dimension (SD) modulation. In multi-dimension (MD) modulation, the users are allowed to modulate both the deblurring and denoising levels.}
		\label{fig:introduction}
	\end{figure}
	\section{Introduction}
	Conventional deep learning methods for image restoration (e.g., image denoising, deblurring and super resolution) learn a deterministic mapping from the degraded image space to the natural image space. For a given input, most of these methods can only generate a fixed output with a pre-determined restoration level. In other words, they lack the flexibility to alter the output effects according to different users' flavors. This flexibility is essential in many image processing applications, such as photo editing, where users desire to adjust the restoration level/strength continuously by a sliding bar. To adapt conventional deep models to real scenarios, several recent works investigate the use of additional branches to tune imagery effects, such as AdaFM~\cite{AdaFM}, CFSNet~\cite{CFSNET}, Dynamic-Net~\cite{dynamic}, DNI~\cite{DNI}, and Decouple-Learning~\cite{fan2018decouple}. The outputs of their networks can be interactively controlled by a single variable at test-time, without retraining on new datasets. They can generate continuous restoration results between the pre-defined “start level” and “end level” (e.g., JPEG quality $q40\rightarrow q10$). 
	
	These pioneer modulation works assume that the input image has only a single degradation type, such as noise or blur, thus the modulation lies in one dimension. However, the real-world scenarios are more complicated than the above assumptions. Specifically, real images usually contain multiple types of degradations, e.g., noise, blur, compression, etc~\cite{yu2018crafting, Suganuma2019CVPR}. Then the users will need separate buttons to control each of them. The solution is far beyond adding more controllable parameters. As these degradations are coupled together, altering a single degradation will introduce new artifacts that do not belong to the pre-defined degradation types. We denote this problem as multi-dimension (MD) modulation for image restoration. Compared with single-dimension (SD) modulation, MD modulation has the following three major differences/difficulties. 
	
	
	\textbf{Joint Modulation.} MD modulation aims to remove the effects of individual degradations as well as their combinations. Different types of degradations are coherently related. Removing one type of degradation could unavoidably affect the other degradations. It is highly challenging to decouple different degradations and modulate each of them separately.
	This can be illustrated in Figure \ref{fig:introduction}. When we only adjust the noise level, the outputs should contain less noise but with fixed deblurring effects. All restored images should also be natural-looking and artifacts-free.
	
	
	\textbf{Zero Starting Point.} In image restoration, the degradation level for modulation can be zero, indicating that the input does not contain the corresponding type of degradation. We call these zero starting points (e,g, $[0, a]$, $[a, 0]$, $[0, 0]$). When the input image has no degradation, restoration algorithm is expected to perform identity mapping. However, this poses challenges for existing SD restoration networks, which usually have information loss in forward processing. Thus it is hard to directly extend current SD methods to the MD task. Please refer to Related Work for details.

	\textbf{Unbalanced Learning.} As there are different degradation types with a large range of degradation levels, the pixel-wise loss (e.g., MSE) will be severely unbalanced for different inputs. For instance, given an input image, the MSE for its blurry version and noisy version could have different orders of magnitude. Furthermore, as the degradation level starts from 0, the MSE can be pretty small around zero points. When we collect these different kinds of data as a training batch, the updating mechanism will favor the patches with large losses and ignore those with small ones. This phenomenon will result in inferior performance on mild degradations. 
	
	To address the aforementioned problems, we propose the first MD modulation framework with dynamic Controllable Residual learning, called CResMD. This is based on a novel use of residual connection. In conventional ResNet~\cite{resnet}, the original input and its residual are combined by direct addition. In our settings, we reformulate it as a weighted sum -- ``$output=input +residual\times \alpha$'', where $\alpha$ is the summation weight. If we add a global residual connection and set $\alpha=0$, the output will be exactly the input. Then we can realize a special case of ``\textit{zero starting point}'' –- identity mapping. 
	In addition, we can also add more local residual connections on building blocks. The underlying assumption is that the building blocks have their unique functions. When we enable some blocks and disable the others, the network can deal with different degradations. Therefore, the ``\textit{joint modulation}'' can also be achieved by dynamically altering the weights $\alpha$. We further propose a condition network that accepts the degradation type/level as inputs and generates the weight $\alpha$ for each controllable residual connection. During training, the base network and the condition network are jointly optimized. 
	To further alleviate ``\textit{unbalanced learning}'', we adopt a new data sampling strategy based on beta distribution. The key idea is to sample more mild degradations than severe ones. 
	
	To verify the effectiveness of the proposed methods,
we conduct extensive experiments on 2D and 3D modulation for deblurring, denoising and JPEG debloking. We have also made comparisons with SD methods (e.g., AdaFM \cite{AdaFM}, CFSNet \cite{CFSNET}, DNI \cite{DNI}) on SD tasks. Experimental results show that the proposed CResMD could realize MD modulation with high accuracy, and achieve superior performance to existing approaches on SD tasks with much less ($0.16\%$) additional parameters

	\section{Related Work}
	\textbf{Image Restoration.} Deep learning methods have been
	widely used in image restoration problems, and most of them focus on a specific restoration task, such as denoising, deblurring, super-resolution and compression artifacts reduction \cite{arcnn, srcnn, dncnn, fsrcnn, srresnet,edsr,ranksrgan}. Here we review some recent works that are designed to handle a wide range of degradation types or levels.
	Zhang et al. \cite{dncnn} propose DnCNN to deal with different levels of Gaussian noise. Then, Guo et al. \cite{cbdnet} attempt to estimate a noise map to improve the denoising performance in real-world applications. Different from these task-specific methods, Yu et al. \cite{yu2018crafting} aim to restore images corrupted with combined distortions with unknown degradation levels by exploiting the effectiveness of reinforcement learning. Later on, they propose a multi-path CNN~\cite{yu2019path} that can dynamically determine the appropriate route for different image regions. In addition, the work in \cite{Suganuma2019CVPR} utilizes the attention mechanism to select the proper operations in different layers based on the input itself. However, these fixed networks cannot be modulated to meet various application requirements.
	
		\textbf{Modulation.} 
	we briefly review four representative SD methods – AdaFM\cite{AdaFM}, CFSNet\cite{CFSNET}, Dynamic-Net\cite{dynamic} and DNI\cite{DNI}. As a common property, all these methods train a couple of networks on two related objectives, and achieve the intermediate results at test time. The main differences lie on the network structure and the modulation strategy. In the first three works, they decompose the model into a main branch and a tuning branch. AdaFM adopts feature modulation filters after each convolution layer. CFSNet uses a side-by-side network upon the main branch and couples their results after each residual block. Dynamic-Net adds modulation blocks directly after some convolution layers. During training, only the tuning branch is optimized to another objective. Due to this finetuning strategy, the modulation could only happen between two objectives. DNI interpolates all network parameters, thus has the flexibility to do MD modulation. However, the linear interpolation strategy of DNI cannot achieve high accuracy (PSNR/SSIM) for image restoration tasks. In contrast, CResMD adopts the joint training strategy with much fewer additional parameters. It could achieve MD as well as SD modulation. 
	
	\begin{figure*}[t]
		\centering
		\includegraphics[scale=0.36]{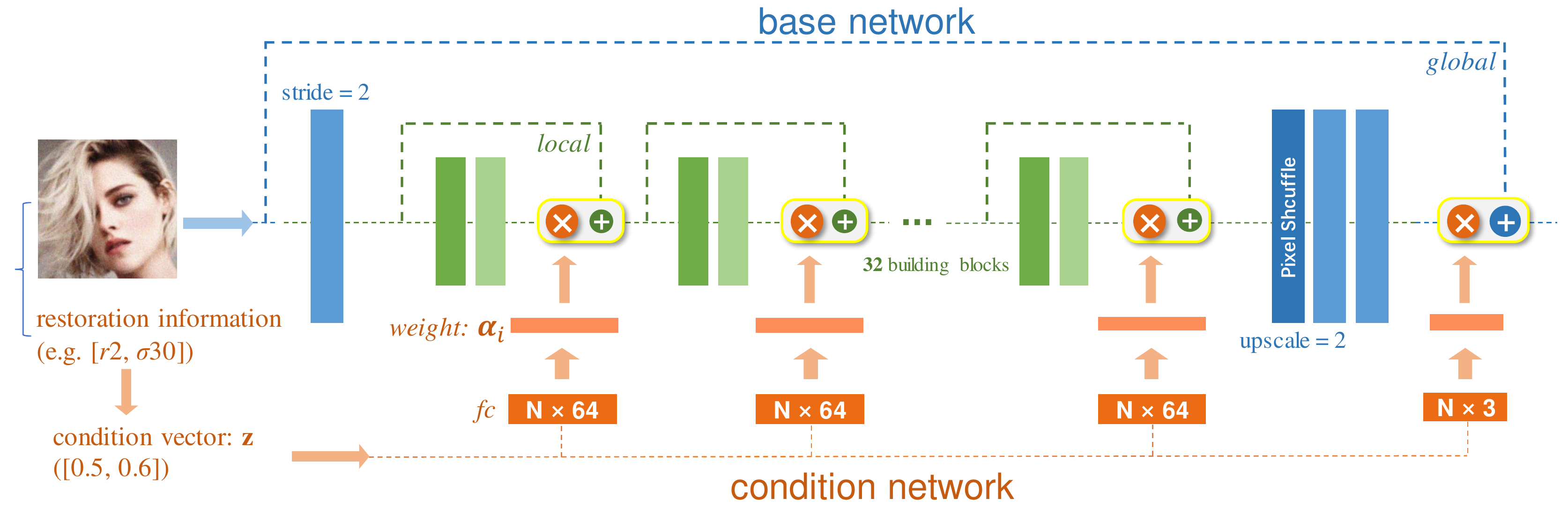}
		\caption{Framework of CResMD, consisting of two branches: base network and condition network. The base network deals with image restoration, while the condition network generates the weights $\alpha$ for the cotrollable residual connections. The condition network contains several fully-connected layers and accepts the normalized restoration information as input. The building block (green) can be replaced by any existing block like residual attention block \cite{Wang_2017_CVPR} or dense block \cite{densenet}.}
		\label{fig:framework}
	\end{figure*}

	\begin{figure*}[t]
	\centering
	\includegraphics[scale=0.3]{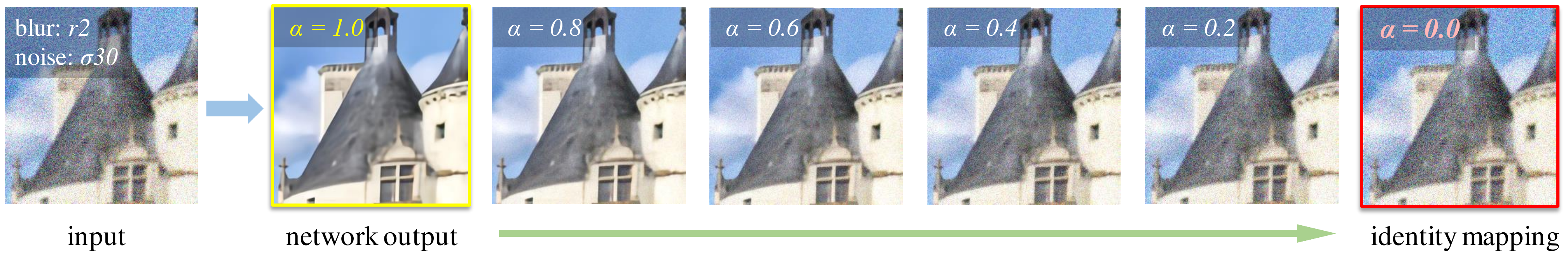}
	\caption{Different levels of restoration effects by setting different weights $\alpha$ on global residual. When $\alpha=1$, the network outputs the restored image. To achieve identity mapping, we set $\alpha=0$ to disable the residual branch.}
	\label{fig:method_global_interpolation}
\end{figure*}

	\section{Method}
	\textbf{Problem Formulation.}
We first give the formulation of multi-dimension (MD) modulation. Suppose there are $N$ degradation types $\{ D_j\}_{j=1}^N$. For each degradation $D_j$, there is a degradation range [0, $R_{j}$]. Our goal is to build a restoration model that accepts the degraded image together with desired restoration information as inputs and generates the restored image. The restoration information (corresponding to the degradation type/level) will act like tool bars, which can be interactively modulated during testing. We use a two-dimension (2D) example to illustrate the modulation process. As shown in Figure \ref{fig:introduction}, there are two separate bars to control the blur level $D_1$ and noise level $D_2$. The modulation space is a square 2D space, spreading from $[0, 0]$ to $[R_{1}, R_{2}]$. We can fix $D_1$ and change $D_2$, then the modulation trajectory is a horizontal line. We can also modulate $D_1$ and $D_2$ simultaneously, then the trajectory will become a diagonal line. If $D_1$ or $D_2$ is fixed on level 0 (zero starting point), then 2D degenerates to 1D. On the contrary, if the starting point is non-zero, such as [0.2, 0.1], then the model cannot deal with [0, 0], [0, $R_2$], [$R_1$, 0]. That is why ``zero starting point'' is essential in MD modulation.

	\textbf{Framework.}
	To achieve MD modulation, we propose a general and effective strategy based on controllable residual connections. The framework is depicted in Figure \ref{fig:framework}. The framework comprises two branches –- a base network and a condition network. The base network is responsible for image restoration, while the condition network controls the restoration type and level. The base network has a general form with downsampling/upsampling layers at two ends and several building blocks in the middle. The building block can be residual block\cite{resnet}, recurrent block\cite{LSTM}, dense block\cite{densenet}, and etc. This structure is widely adopted in advanced image restoration models \cite{srresnet,edsr,ResDense,dncnn,cbdnet}. The only difference comes from the additional ``controllable residual connections'', shown as blue and green dash lines in Figure \ref{fig:framework}. These residual connections are controlled by the condition network. Take any degradation type/level as input, the condition network will first convert them into a condition vector, then generate the weights $\alpha$ for controllable residual connections. At inference time, we can modulate the degradation level/type -- $\{D_i\}_{i=1}^N$, then the model can generate continuous restoration results.

	\textbf{Controllable Residual Connection.}
	The proposed controllable residual connection comes from the standard residual connection, thus it is essential to review the general form of residual connection. Denote $X$ and $Y$ as the input and output feature maps. Then the residual connection can be represented as
	\begin{align}
	\textit{Y} = f(\textit{X}, {W_{i}}) + \textit{X},
	\end{align}
	where $f(\textit{X}, {W_{i}})$ refers to the residual feature maps and $f(\cdot)$ is the mapping function. While in our controllable residual connection, we add a tunable variable $\alpha$ to control the summation weight. The formulation becomes
	\begin{align}
	\textit{Y} = f(\textit{X}, {W_{i}})\times \alpha + \textit{X},
	\end{align}
	where $\alpha$ has the same dimension as the number of feature maps. This simple change gives residual connection two different properties. First, through tuning the variable $\alpha$ from 0 to 1, the output $Y$ will change continuously from $X$ to $f(X, W_i)+X$. Second, the residual part can be fairly skipped by setting $\alpha=0$. We can add the following two types of controllable residual connections.

	(1) Global connection -- $X,Y$ are input/output images. The initial motivation of adding global connection is to handle the extreme case of zero starting point, where all degradation levels are zero. Generally, it is hard for a conventional neural network to perform identity mapping and image restoration simultaneously. However, with the help of global connection, the identity mapping can be easily realized by setting $\alpha=0$. Furthermore, when we change the values of $\alpha$, the output will exhibit different levels of restoration effects. This phenomenon is illustrated in Figure \ref{fig:method_global_interpolation}, where the input image is degraded by noise+blur and the intermediate results are obtained by using different $\alpha$. 
	
	(2) Local connection -- $X,Y$ are input/output feature maps. If the imagery effects can be affected by a simple variable, we can also control the feature maps to achieve more complicated transformation. A reasonable idea is to add local residual connection on each function unit, which is responsible for specific degradation. By disabling/suppressing some function units, we can deal with different degradations. However, it is almost impossible to decouple these degradations and define a clear function for each block. Thus we roughly group some basic building blocks and add controllable residual connections. The minimum function unit consists of a single building block. Experiments in Figure \ref{fig:curve_skip} show that more local residual connections achieve better performance at the cost of more controlling variables. More analyses can be found in Section 4.4.

	\textbf{Condition Network.}
	We further propose a condition network that accepts the degradation type/level as input and generates the weight $\alpha$ for each controllable residual connection. As each degradation has its own range, we should first encode the degradation information into a condition vector. Specifically, we linearly map each degradation level to a value within the range [0, 1], and concatenate those values to a vector $\textit{\textbf{z}}$. Then the condition vector is passed through the condition network, which can be a stack of fully-connected layers (see Figure~\ref{fig:framework}). 
\textbf{Data Sampling Strategy.}
Data sampling is an important issue for MD modulation. As the training images contain various degradation types/levels, the training loss will be severely biased. If we uniformly sample these data, then the optimization will easily ignore the patches with small MSE values, and the performance of mild degradations cannot be guaranteed. To alleviate the unbalanced learning problem, we sample the degradation levels for each degradation type based on the beta distribution:
	\begin{align}
	g(z, \alpha, \beta) = \frac{1}{B(\alpha, \beta)}z^{\alpha-1}(1-z)^{\beta-1}.
	\label{func:beta}
	\end{align}
	
\begin{figure}[htbp]
	\centering
	\includegraphics[scale=0.33]{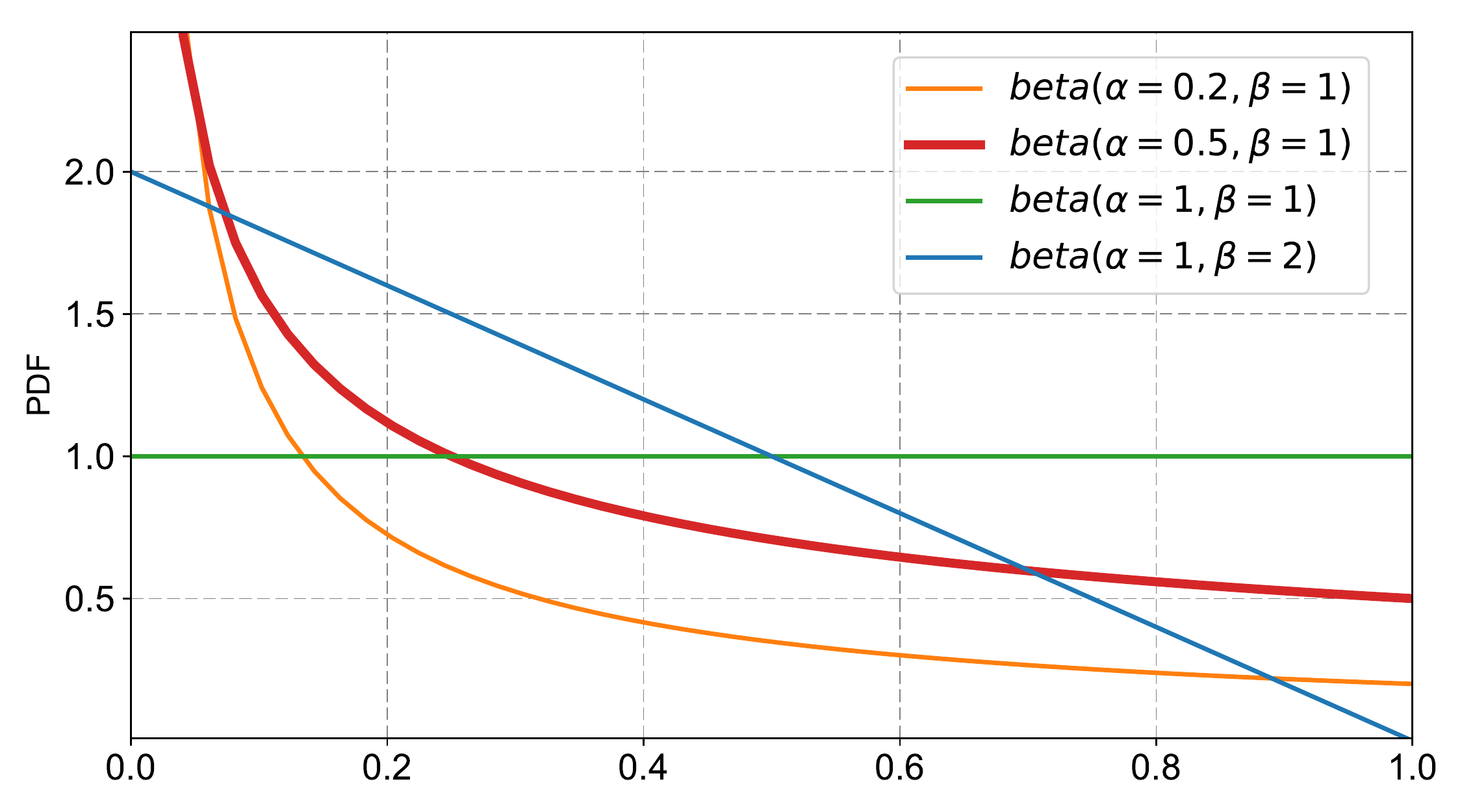}
	\caption{Beta distribution.}
	\label{fig:beta_dis}
\end{figure}

As shown in Figure \ref{fig:beta_dis}, a larger value of $\alpha$ is associated with a steeper curve, indicating that the sampled degradation levels are inclined to the mild degradations. In our experiments, $\alpha$ and $\beta$ are set to 0.5 and 1, respectively. We also compare the results of different sampling curves in Section 4.4.
	
	
	\textbf{Training and Testing.}
	The training strategy is straightforward. We generate the training data with different degradations and their combinations by random sampling, and encode the degradation information into the condition vector (range [0, 1]). Note that the training data is artificially generated, thus the degradation information is known during training. The model takes both the corrupted image and the condition vector as inputs. The original clean image is used as ground truth. The joint training based on L1 loss will enable different restoration effects under different condition vectors.
	
	In the testing stage, the degradation information is unknown, thus the users can modify the elements of the condition vector to obtain various restoration results. In other words, the condition vector refers to the restoration strength and performs like sliding bars. For example, given a corrupted image with blur level $r =2$ (range [0, 4]) and noise level $\sigma=30$ (range [0, 50]), the users are free to modulate the denoising/debluring bars (condition vector) with any sequence or simultaneously, and finally find the best choice at around [2/4, 30/50].
	
	\textbf{Discussion.} The proposed CResMD is a simple yet effective method that is specially designed for MD modulation. In comparison, existing SD methods cannot be directly extended to the MD task, mainly for two reasons. (1) The training strategy determines that their modulation trajectory cannot span across the 2D space. Specifically, their models generally have a main branch and a tuning branch. The main brunch is trained for the first objective, and the tuning brunch is fine-tuned on another objective. Thus the modulation trajectory is a line connecting two objectives. Even we use multiple parameters, the modulation trajectory will become a diagonal line instead of a 2D space. For example, if the start level is $[0.1, 0.1]$ and the end level is $[1, 1]$, the model can deal with $[0.5, 0.5]$ but not $[0.4, 0.6]$.  In order to achieve joint modulation, we cannot just use degradations on the start and end levels, but should consider all combinations of degradations between two ends. (2) The network structure prevents them from realizing “zero starting point”. For example, if the start level is $[0,0]$, the main branch will perform identity mapping (output=input). Then it is hard to adapt this network to another objective only by modulating intermediate features. That is why we propose to use skip/residual connection with controllable parameters.
	

	\section{Experiments}
	\subsection{Implementation Details}
	We first describe the network architectures. For the base network, we adopt the standard residual block as the building block, which consists of two convolution layers and a ReLU activation layer. There are 32 building blocks, of which each convolution layers have 64 filters with kernel size $3\times3$. In order to save computation, we use strided convolution to downsample the features to half size. The last upsampling module uses a pixel-shuffle \cite{pixelshuffle} layer followed by two convolution layers. Note that the first and last convolution layers are not followed by ReLU activation. We add a local controllable residual connection on each building block. For the condition network, we use a single fully-connected layer to output a 64-dimension vector $\alpha$ for each local controllable residual connection. In total, there are 32 layers for 32 local connections and 1 layer for the global connection. 
	
	To ease the burden of evaluation, we conduct most experiments and ablation studies on 2D modulation. To demonstrate the generalization ability, we conduct an additional experiment on 3D modulation at last. In 2D experiments, we adopt two widely-used degradation types – Gaussian blur and Gaussian noise. JPEG compression is further added in the 3D experiment. 
	
	The training dataset is DIV2K\cite{DIV2K}, and the test datasets are CBSD68 \cite{CBSD68} and LIVE1 \cite{LIVE1}. The training images are cropped into $64\times 64$ sub-images. To generate corrupted input images, we employ mixed distortions on the training data. In particular, blur, noise and JPEG are sequentially added to the training images with random levels. For Gaussian blur, the range of kernel width is set to $r\in [0, 4]$, and the kernel size is fixed to $21 \times 21$. The covariance range of Gaussian noise is $\sigma\in [0, 50]$, and the quality range of JPEG compression is  $q\in [100, 10]$. We sample the degradations with stride of 0.1, 1, and 2 for blur, noise, and JPEG compression, respectively.
	
	These training images are further divided into two groups, one with individual degradations and the other with degradation combinations. To augment the training data, we perform horizontal flipping and 90-degree rotation. To obtain more images with mild degradations, we force the sampling to obey beta distribution, where $\alpha$ and $\beta$ are set to 0.5 and 1, respectively. The mini-batch size is set to 16. The L1 loss is adopted as the loss function. During the training process, the learning rate is initialized as  $5 \times 10^{-4}$, and is decayed by a factor of 2 after $2\times 10^{5}$ iterations.  All experiments run $1 \times 10^{6}$ iterations. We use PyTorch framework and train all models on NVIDIA 1080Ti GPUs.
	
	\subsection{Complexity Analysis}
	
	The proposed CResMD is extremely light-weight, contributing to less than 4.2k parameters. As the additional parameters come from the condition network, the number of introduced parameters in 2D modulation is calculated as $32\times 2\times 64 + 2\times 3 = 4102$. Note that the base network contains 32 building blocks with parameters around 2.5M, CResMD only comprises 0.16\% of entire model. In contrast, the tuning blocks in AdaFM and CFSNet account for 4\% and 110\% of the total parameters of the base network, respectively. 
	Another appealing property is that the computation cost of condition network is a constant, as there are no spatial or convolution operations. In other words, the computation burden is nearly negligible for a large input image.  
	
		\subsection{Performance Evaluation}
	To evaluate the modulation performance, we follow AdaFM \cite{AdaFM} and use PSNR distance. Specifically, if we want to evaluate the performance on ${D_1,D_2}$, then we train a baseline model using the architecture of the base network on ${D_1,D_2}$. This baseline model can be regarded as an upper bound. With the ground truth images, we can calculate PSNR of CResMD and the baseline model respectively. Their PSNR distance is used as the evaluation metric (lower is better).
	
	\textbf{2D modulation.}
	First, we evaluate the 2D modulation performance of the proposed method. The quantitative results\footnote{Results on more datasets can be found in supplementary file.} of different degradations on CBSD68 dataset are provided in Table \ref{Table:blur_noise_diff}. 
	We can observe different trends for different degradation types. For two degradations, the PSNR distances are all below 0.2 dB, indicating a high modulation accuracy. For one degradation, where there are zero starting points, the performance will slightly decrease. Furthermore, blur generally leads to higher PSNR distances than noise. The largest PSNR distance appears in $r1$, which is a starting point as well as a mild degradation. Nevertheless, its absolute PSNR value is more than 38 dB, thus the restoration quality is still acceptable. We further show qualitative results in Figure \ref{fig:md_modulation}, where all images exhibit smooth transition effects. 

	\begin{table*}[t]
		\centering
		\renewcommand{\arraystretch}{1}
		\setlength{\tabcolsep}{1.2pt}
		\caption{2D experiments evaluated on CBSD68 \cite{CBSD68}. The PSNR distances within 0.2 dB are shown in bold. Lower is better.}
		\begin{tabular}{@{}rcccccccccccccc@{}}
			
			\toprule
			& \multicolumn{6}{c}{$\textit{\textbf{one degradation}}$} & & \multicolumn{6}{c}{$\textit{\textbf{two degradations}}$} &
			\\
			\cmidrule{2-7} \cmidrule{9-14}
			
			blur $r$& 1& 2 & 4 & 0 & 0 & 0 &&  1 &1  &2&2&4&4\\
			
			noise $\sigma$& 0 & 0& 0& 15 & 30 & 50  &&15  &50 &15&50 &15  &50\\ \midrule
			upper bound & 39.07 & 30.24 & 26.91 & 34.12 & 30.56 & 28.21 && 29.11  & 26.07 & 26.30 & 24.55 &24.08&  23.03\\
			
			CResMD & 38.38 & 30.09 & 26.53 & 33.97 & 30.43 & 28.06 && 29.00 & 25.96 & 26.24 & 24.48 & 24.03  & 22.95\\
			\cmidrule{2-14}
			PSNR distance  & 0.69& \textbf{0.15}& 0.38 & \textbf{0.15}& \textbf{0.13}& \textbf{0.15} && \textbf{0.11}  &\textbf{0.11}&\textbf{0.06}
			&\textbf{0.07}&\textbf{0.05} &\textbf{0.08}\\
			%
			\midrule
		\end{tabular}
		\label{Table:blur_noise_diff}
	\end{table*}

	\textbf{Comparison with SD methods.}
As the state-of-the-art methods are all proposed for SD modulation, we can only compare with them on single degradation types. We want to show that even trained for MD modulation, CResMD can still achieve excellent performance on all SD tasks. Specifically, we compare with DNI, AdaFM, and CFSNet on deblurring, starting from $r2$ to $r4$. Deblurring is harder than denoising, thus could show more apparent differences. To re-implement their models, we first train a base network on the start level $r2$. Then we finetune (1) the whole network in DNI, (2) the AdaFM layers with kernel size $5\times5$ in AdaFM, (3) the parallel tuning blocks and coefficient network in CFSNet, to the end level $r4$. To obtain the deblurring results between $r2$ and $r4$, we interpolate the networks of two ends with stride 0.01. For CResMD, we directly use the deblurring results in the 2D experiments (Table~\ref{Table:blur_noise_diff}). 
\begin{figure}[htbp]
	\centering
	\includegraphics[scale=0.45]{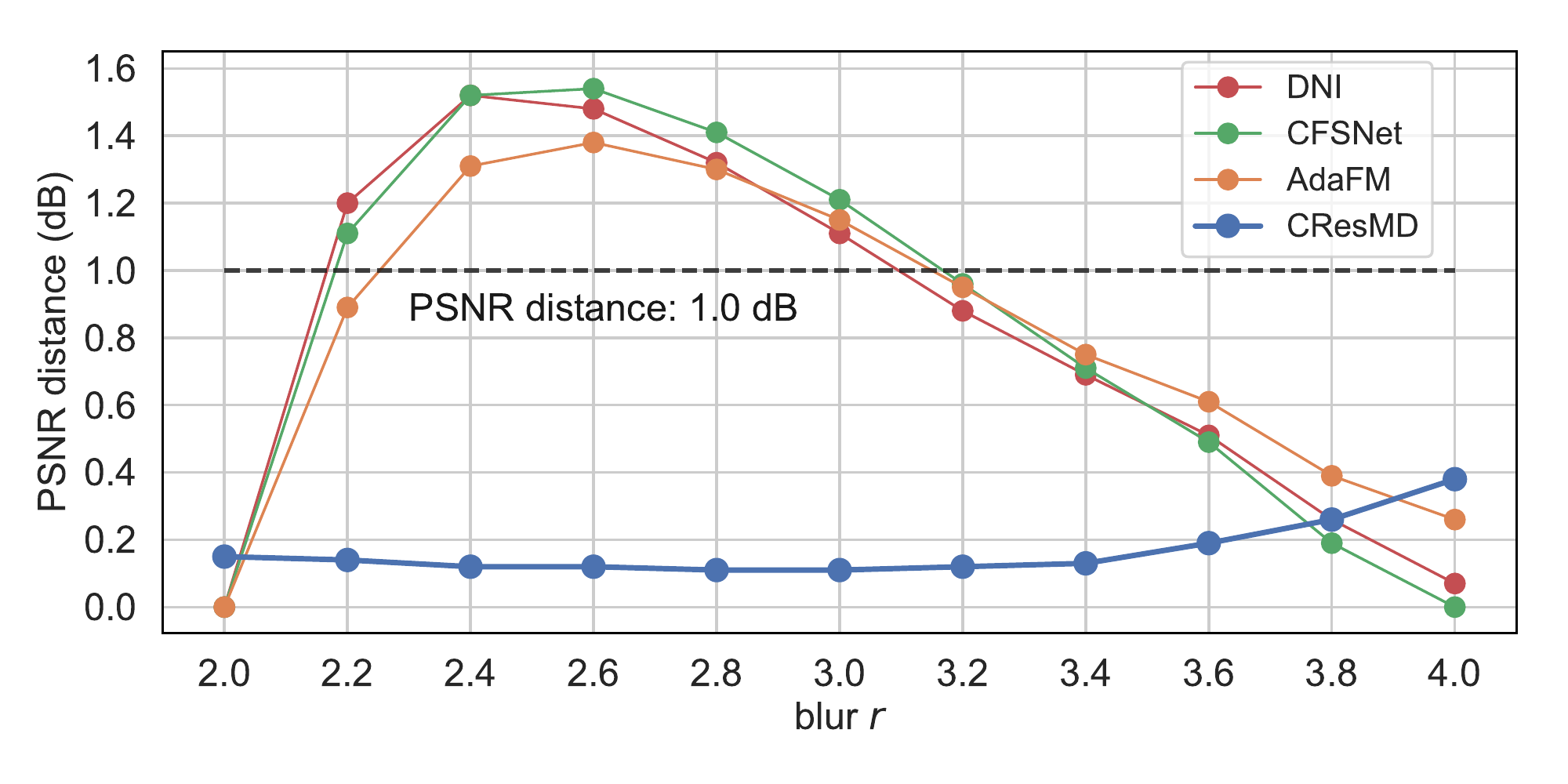}
	\caption{Quantitative comparison with SD methods on CBSD68 data set in PSNR.}
	\label{fig:DNI_CFSNet_adafm}
\end{figure}

From Figure \ref{fig:DNI_CFSNet_adafm}, we observe that our method significantly outperforms the others in almost all intermediate points. In particular, the SD methods tend to yield high PSNR distances ($>1.0$ dB) on tasks $r2.2 \sim r3.0$. It is not surprising that they perform perfect at two ends as they are trained and finetuned on these points. This trend also holds for denoising, but with much smaller distances. All these results demonstrate the effectiveness of our proposed method in SD modulation. Results of deblurring $r1 \rightarrow r4$, denoising $\sigma5 \rightarrow \sigma50$ and $\sigma15 \rightarrow \sigma50$ can be found in the supplementary file.


	%
	%
	%

\subsection{Ablation Study}

\textbf{Effectiveness of Global Connection.}
The global connection is initially designed to handle the problem of zero starting point. In general, it is hard for a conventional network to deal with both identity mapping and image restoration at the same time. With the proposed controllable global connection, we can ideally turn off the residual branch by setting $\alpha=0.$ To evaluate its effectiveness, we conduct a straightforward comparison experiment by just removing the global connection. This new model is trained under the same setting as CResMD. As for testing, we only select those mild degradations, such as blur $r<1$ and noise $\sigma<15$. It is clear that the model with controllable global connection could achieve better performance on all mild degradations as we can see from Table~\ref{table:global_residual}.
	\begin{table}[htbp]	
		\centering
		\setlength{\tabcolsep}{3pt}
		\renewcommand{\arraystretch}{1}
		\caption{The effectiveness of global connection. Results are evaluated by PSNR.}
		\label{table:global_residual}
		\begin{tabular}{@{}rcccccccc@{}}
			\toprule
			
			
			blur $r$&0& 0 &  0.5 & 1 & 0.5 & 0.5 & 1 \\ 
			noise $\sigma$& 0& 5 &  0 & 0 & 5 & 15 & 5 \\
			\midrule
			CBSD68 \,w/o &71.39 & 40.21 & 52.70 & 38.04 & 37.80 & 32.31 &31.48\\
			w & $+\infty$ & 40.33 & 53.17 & 38.38 &37.92&32.44&31.63\\
			
			\midrule
			gain &$+\infty$ & $0.12$ & $0.47$ & $0.34$ &$0.12$&$0.13$&$0.15$\\
			\midrule
			LIVE1 \,w/o &64.17 & 39.79 & 51.22 & 38.38  &37.71 & 32.51 & 31.65 \\
			w &$+\infty$ & 39.99 & 52.21 & 38.85 & 37.89 & 32.69 & 31.86 \\
			\midrule
			gain &$+\infty$ & $0.20$  &$0.99$ & $0.47$ &$0.18$&$0.18$&$0.21$\\
			\bottomrule
		\end{tabular}
	\end{table}
	
	
	%
\begin{figure*}[t]
	\begin{minipage}[]{0.33\linewidth}
		\flushright
		\subfigure[$1\; group$]{\includegraphics[scale=0.4]{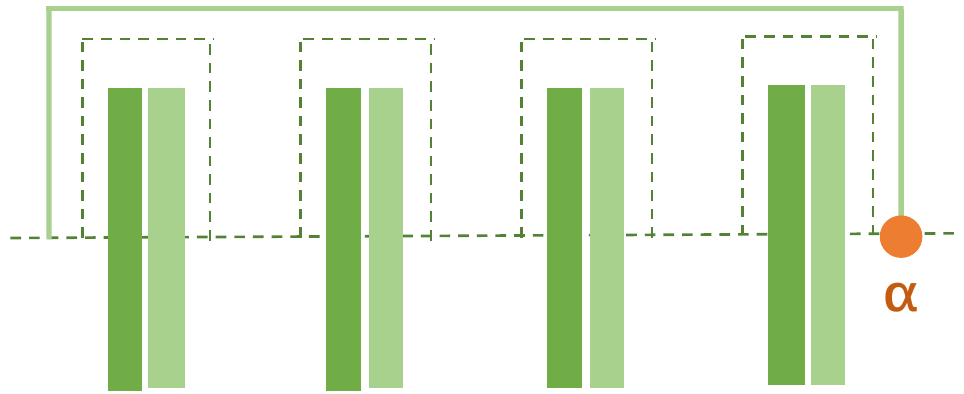}}
	\end{minipage}%
	\begin{minipage}[]{0.33\linewidth}
		\centering
		\subfigure[$2\; groups$]{\includegraphics[scale=0.39]{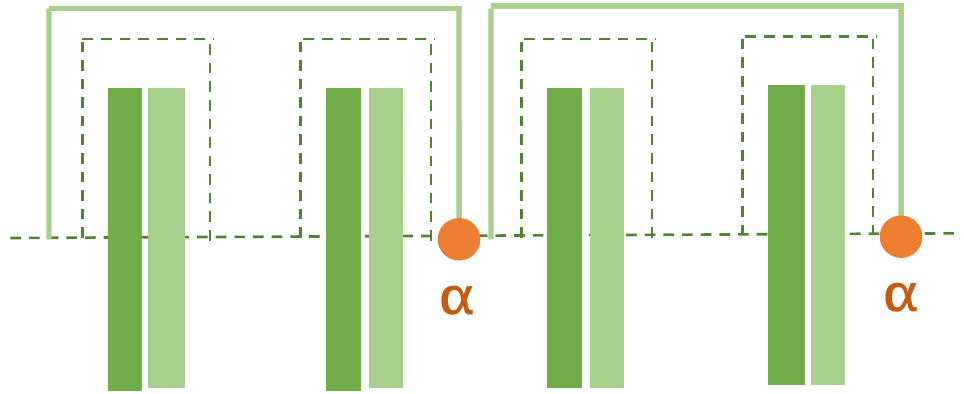}}
	\end{minipage}
	\begin{minipage}[]{0.33\linewidth}
		\flushleft
		\subfigure[$4\; groups$]{\includegraphics[scale=0.41]{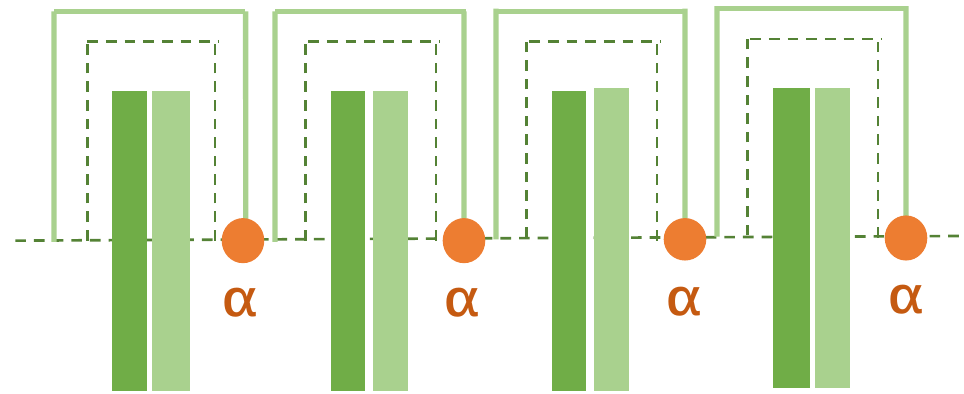}}
	\end{minipage}
	\caption{Different options of local connections.}
	\label{figure_skip}
\end{figure*}

\begin{figure*}[!h]
	\includegraphics[scale=0.24]{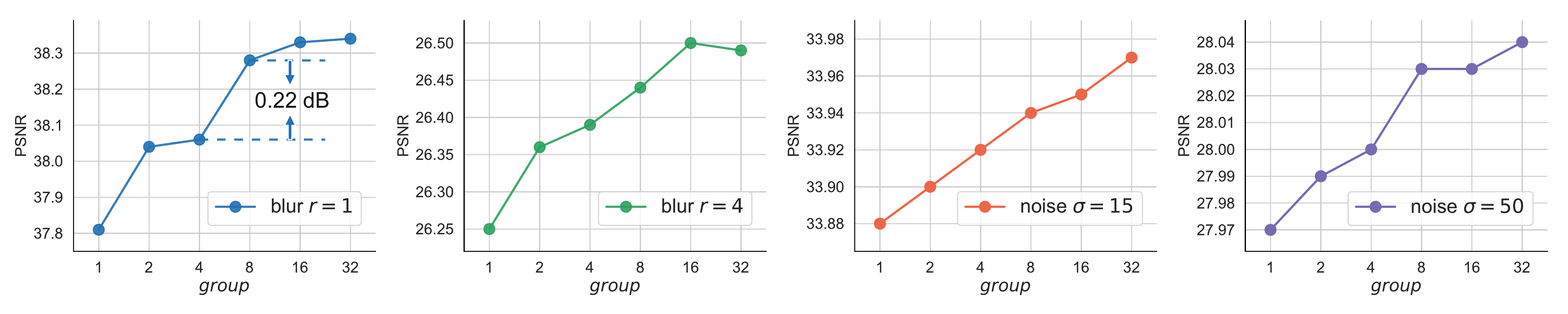}
	\caption{Performance under different local connections evaluated by CBSD68 dataset.}
	\label{fig:curve_skip}
\end{figure*}

	\textbf{Effectiveness of Local Connection.}
Here we test the influence of the number of local connections. In particular, we group some basic building blocks as a function unit and add controllable residual connection. All the building blocks are divided into 1, 2, 4, 8, 16 and 32 groups (the details are illustrated in Figure \ref{figure_skip}). They are evaluated in 2D modulation on CBSD68 dataset. The results are depicted in Figure~\ref{fig:curve_skip}. Obviously, more groups or local connections could lead to better performance. Particularly, we also observe a sharp leap (0.22dB) in deblurring $r1$ (from 4 to 8 local connections), indicating that at least 8 local connections are required. In contrast, results on denoising tasks are less significant, where the PSNR distance between 1 and 32 local connections is less than 0.1dB in denoising $\sigma15$ and $\sigma50$.

	\textbf{Effectiveness of Data Sampling.} 
After analysis of the proposed network structures, we then investigate different data sampling strategies. As mentioned in Section 3.4, appropriate data sampling strategies could help alleviate the unbalanced learning problem. To validate this comment, we conduct a set of controlled experiments with different sampling curves, which can be generated using different parameters of beta distribution in Function~\ref{func:beta}. To be specific, the most commonly used strategy is uniform sampling, corresponding to the green horizontal line in Figure~\ref{fig:beta_dis}. We can generate this curve by setting $\beta$ and $\alpha$ to 1. 
Similarly, we can further set $\alpha, \beta$ to be (0.5, 1.0), (0.2, 1.0) and (1.0, 2.0) to generate linear and non-linear curves, shown in Figure~\ref{fig:beta_dis}. Then we train four CResMDs on different training datasets with the above sampling strategies. Results are shown in Table~\ref{loss_weight}, where we use uniform sampling ($\alpha=1,\beta=1$ ) as our baseline and calculate the PSNR distances with other strategies. Obviously, when we sample more data on mild degradations, the performance will significantly improve. Furthermore, the PSNR increases on some degradation levels generally comes at the cost of the decrease on the others. For instance, in deblurring $r=1$, $\alpha=1.0, \beta=2.0$ and $\alpha=0.2, \beta=1.0$ reach the highest performance, but also get severe degradation in $r4$.  As a better trade-off, we select the setting $\alpha=0.5$, $\beta=1.0$ for our CResMD, which stably improves most degradation levels.

	\begin{table}[t]
	\centering
	\setlength{\tabcolsep}{0.5pt}
	\renewcommand{\arraystretch}{1}
	\caption{Performance under different sampling curves evaluated on LIVE1 \cite{LIVE1}. results are given in PSNR.}
	\label{loss_weight}
	\begin{tabular}{@{}rcccccccccccccc@{}}
		\toprule
		
		blur $r$&& 1 & 2 & 4 && 0 & 0 & 0 && 1 & 2 & 4& \\ 
		noise $\sigma$&& 0 & 0 & 0 && 5 & 30 & 50 && 5 & 30 & 50& total&\\
		\midrule
		\quad$\alpha=1.0,\beta=1.0$ && 38.66 & 30.01 &\textbf{26.26} && 39.90 & 30.63 & 28.24 &&31.78&24.97 & \textbf{22.58}&&\\
		\midrule
		$\alpha=0.5,\beta=1.0$ &&38.85 & 30.03 &26.14 && 39.99 &30.65&28.25&& 31.86 &24.98&22.56&&\\
		
		(CResMD)&& $+0.19$ & $+0.02$ & $-0.12$ && $+0.09$ &$+0.02$&$\textbf{+0.01}$&&$+0.08$&$+0.01$&$-0.02$&$\textbf{+0.28}$& \quad\\
		\midrule
		
		
		$\alpha=0.2,\beta=1.0$ &&38.94 & 29.98 &26.07 && 39.97 &30.55&28.10&&31.68&24.85&22.39&&\\
		
		&& $\textbf{+0.28}$ & $-0.03$ & $-0.19$ && $+0.07$ &$-0.08$&$-0.14$&&$-0.10$&$-0.12$&$-0.19$&$-0.50$&\\
		\midrule
		
		$\alpha=1.0,\beta=2.0$&& 38.93 & 30.08 &25.80 && 40.00&30.66&28.24&&31.90&24.99&22.50&&\\
		
		&& $+0.27$ & $\textbf{+0.07}$ & $-0.46$ && $\textbf{+0.10}$ &$\textbf{+0.03}$&$+0.00$&&$\textbf{+0.12}$&$\textbf{+0.02}$&$-0.08$&$+0.07$&\\
		
		%
		%
		\bottomrule
	\end{tabular}
\end{table}

	\begin{figure*}[t]
	\begin{minipage}[]{0.9\linewidth}
		\centering
		\subfigure[2D modulation.]{\includegraphics[scale=0.24]{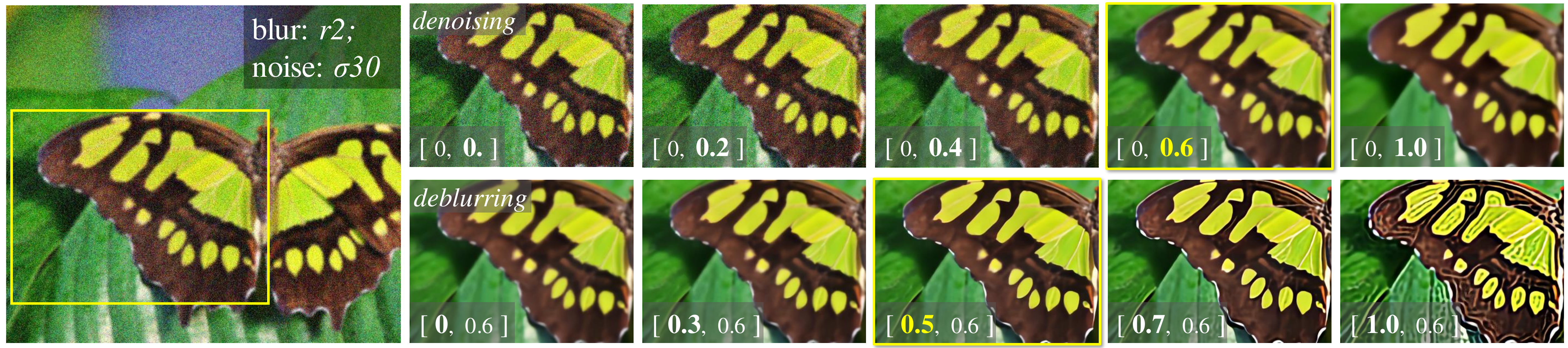}}
		
	\end{minipage}%
	
	\begin{minipage}[]{0.9\linewidth}
		\centering
		\subfigure[3D modulation.]{\includegraphics[scale=0.293]{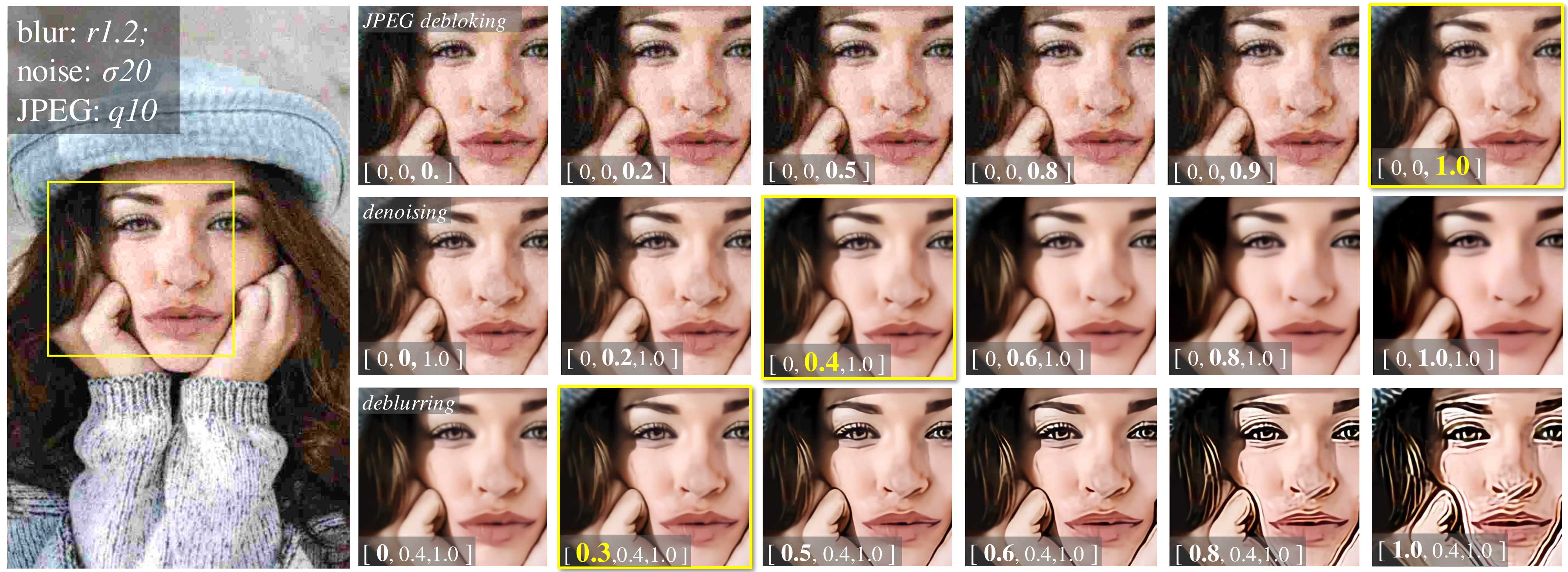}}
		
	\end{minipage}
	\caption{Qualitative results of MD modulation. In each row, we only change one factor with other factors fixed. We arrive at the best choice in the yellow box. \textbf{Better view in zoom and color.} }
	\label{fig:md_modulation}
\end{figure*}


	\textbf{Generalization to 3D modulation.}
	In the above experiments, we mainly use 2D modulation for illustration. Our method can be easily extended to higher dimension cases. Here we show a 3D modulation example with three degradation types: blur, noise and JPEG compression. Note that in JPEG compression, the zero starting point is not quality $100$ but quality $\infty$, thus we extend the JPEG range as $\{\infty, [100, 10]\}$. In 2D modulation, there is only one degradation combination – noise+blur. However, in 3D, the number increases to 4, including noise+blur, noise+JPEG, blur+JPEG and noise+blur+JPEG. Then the difficulty also improves dramatically. Nevertheless, our method can handle this situation by simply setting the dimension of the condition vector to 3. All the other network structures and training strategy remain the same. 
	The results can be found in the supplementary file.
	We can observe that most PSNR distances are below 0.3 dB, indicating a good modulation accuracy. 
	Compared with 2D modulation, the performance on single degradations decreases a little bit, which is mainly due to the insufficient training data. We also show some qualitative results in Figure~\ref{fig:md_modulation}, where we modulate one factor and fix the others. 
	
	 Experimentally, more degradation types ($\textgreater3$) require larger datasets/networks, and the performance on individual tasks will also degrade. Considering that there are not so many degradation types in real-world scenarios, our method can deal with most cases of image restoration tasks. We leave the higher dimension problem to future research.

	\section{Conclusion}
	In this work, we first present the multi-dimension modulation problem for image restoration, and propose an efficient framework based on dynamic controllable residual learning. With a light-weight structure, the proposed CResMD partially addresses the three difficult problems in MD modulation. Although CResMD could realize modulation across multiple domains, the performance can be further improved. The controlling method can be more accurate and diverse. We encourage future research on better solutions. 
	
\noindent\textbf{Acknowledgement}
This work is partially supported by the National Natural Science Foundation of China (61906184), Science and Technology Service Network Initiative of Chinese Academy of Sciences (KFJ-STS-QYZX-092), Shenzhen Basic Research Program (JSGG20180507182100698, CXB201104220032A), the Joint Lab of CAS-HK，Shenzhen Institute of Artificial Intelligence and Robotics for Society.

%
%

\bibliographystyle{splncs04}

\end{document}